\documentclass{PoS}

\title{Anomalous Chiral Transport in Heavy Ion Collisions}

\ShortTitle{Anomalous Chiral Transport in Heavy Ion Collisions}

\author{\speaker{Jinfeng Liao}
\thanks{Research supported in part by the U.S. National Science Foundation (Grant No. PHY-1352368) and by the U.S. Department of Energy, Office of Science, Office of Nuclear Physics, within the framework of the Beam Energy Scan Theory (BEST) Topical Collaboration. The author thanks X. Huang, Y. Jiang, E. Lilleskov, S. Shi and Y. Yin for collaborations on various aspects of the results reported in this contribution.}\\
        Physics Department and Center for Exploration of Energy and Matter,
Indiana University,\\ 2401 N Milo B. Sampson Lane, Bloomington, IN 47408, USA.\\
        E-mail: \email{liaoji@indiana.edu}}

\abstract{Chiral anomaly is a very fundamental aspect of  quantum theories with chiral fermion, from the Standard Model to supersymmetric field theories or even string theories. How such microscopic anomaly manifests itself in a macroscopic many-body system with chiral fermions, is a highly nontrivial question that has recently attracted significant interest. As it turns out, unusual transport currents can be induced by chiral anomaly under suitable conditions in such systems, with the notable example of the Chiral Magnetic Effect (CME) where a vector current (e.g. electric current) is generated along an external magnetic field. The CME has been enthusiastically studied in two very different physical systems: the Dirac and Weyl semimetals in condensed matter physics as well as the quark-gluon plasma in heavy ion collisions. In this contribution, we report the latest theoretical and experimental status for the search of CME in heavy ion collisions.}

\FullConference{38th International Conference on High Energy Physics\\
		3-10 August 2016\\
		Chicago, USA}

\begin{document}

\section{Introduction}

Symmetry principles play instrumental roles in the construction of our most basic physical theories. A special category of ``symmetry'' is the so-called anomaly, which is a well-defined classical symmetry of a theory but gets broken at quantum level. A most famous example of anomaly is the chiral anomaly  which is a very fundamental aspect of  quantum theories with spin-$\frac{1}{2}$ chiral fermion, from the Standard Model to supersymmetric field theories or even string theories. In such theories the right-handed(RH) or left-handed(LH) chiral current $J^\mu_{R/L}$, while classically conserved, is no longer conserved once coupled to gauge fields quantum mechanically:  
\begin{eqnarray} \label{eq_CA}
\partial_\mu J^\mu_{R/L} = \pm C_A   E_\mu B^\mu 
\end{eqnarray}
where $C_A$ is a universal coefficient and the gauge fields $E,B$ could be Abelian or non-Abelian. In the context of strong interaction physics as described by Quantum Chromodynamics (QCD), the chiral anomaly provides a unique access to the topological configurations such as instantons and sphelarons. These objects are deeply associated with the origin of the most salient nonperturbative phenomena like confinement and spontaneous chiral symmetry breaking in QCD~\cite{Schafer:1996wv,Diakonov:2009jq}. By virtue of (\ref{eq_CA}) the topological fluctuations via such objects could be ``translated'' into the chirality fluctuations of fermions in the system which would then be experimentally accessible.

Microscopic symmetry principles also manifest themselves nontrivially in macroscopic physics. For example, time translation invariance and global phase invariance imply energy and charge conservation respectively, which further  necessitate temperature and chemical potential as proper thermodynamic variables for describing macroscopic systems. Hydrodynamics, as the long-time large-distance effective description of any macroscopic system, is an even more direct manifestation of symmetries. The hydrodynamic equations for energy-momentum tensor and for charged current are the direct consequences of conservation of energy, momentum and charge which all originate from corresponding microscopic symmetries.  

Recently there has been significant interest in understanding the implication of microscopic quantum anomaly (as a sort of ``half symmetry'') on the macroscopic properties of matter. As it turns out, unusual transport currents can be induced by chiral anomaly under suitable conditions in such systems, with the notable example of the Chiral Magnetic Effect (CME) where a vector current (e.g. electric current) is generated along an external magnetic field $B^\mu$~\cite{Kharzeev:2004ey,Fukushima:2008xe}:
\begin{eqnarray} \label{eq_CME}
J^\mu  =  C_A   \mu_5 B^\mu 
\end{eqnarray}
where the coefficient $C_A$ is totally dictated by anomaly relation in (\ref{eq_CA}) and the $\mu_5$ is a chiral chemical potential that quantifies the macroscopic imbalance between RH and LH fermions. Remarkably, the   anomalous transport process underlying the above equation is time-reverse even, i.e. non-dissipative.  Given the significance of this phenomenon, it is of great interest to search for its experimental signal in real materials. 
The CME has been enthusiastically studied in two very different physical systems: the Dirac and Weyl semimetals in condensed matter physics~\cite{Li:2014bha} as well as the quark-gluon plasma in heavy ion collisions~\cite{STAR_LPV1,STAR_LPV_BES,ALICE_LPV}. See recent reviews in e.g. \cite{Kharzeev:2015znc,Liao:2014ava,Huang:2015oca}. In this contribution, we report the latest theoretical developments and   experimental measurements  for the search of CME in heavy ion collisions. We also briefly discuss  other interesting anomalous transport effects induced by chiral anomaly.

\section{Anomalous-Viscous Fluid Dynamics (AVFD)}

In heavy ion collisions, two large nuclei collide at very high energy and create a zone of extremely high energy density, where a quark-gluon plasma (QGP) at a temperature about a few hundred MeV (or, of the order of about trillion degrees) forms. This QGP will expand violently outward and thus cools down for a transient period of time before its eventual freeze-out into hadrons. Such collision experiments are now carried out at the Relativistic Heavy Ion Collider (RHIC) and the Large Hadron Collider (LHC). From extensive studies at RHIC and the LHC, it has been found that the QGP is a nearly perfect fluid with extremely small shear viscosity, and  its bulk expansion is accurately described by relativistic viscous fluid dynamics. 

Such a QGP has (nearly) massless quarks (i.e. up and down flavors) as chiral fermions, and our interest is to look for the anomalous transport such as CME (\ref{eq_CME}) arising from the chiral anomaly associated with these quark currents. As already said, the QGP in heavy ion collisions flows hydrodynamically. This naturally raises the question: how should one modify the usual hydrodynamic equations for exactly conserved currents into a new proper description of currents that bear anomaly? The question was answered in a very nice paper \cite{Son:2009tf}: the fluid dynamic equation for an anomalous current takes the form of (\ref{eq_CA}) and more importantly the constituent relation for such  fermion currents (at first order of gradient expansion) is required by the second law of thermal dynamics to include anomalous terms corresponding to the CME term (\ref{eq_CME}) and a similar chiral vortical current. A number of initial attempts were made to apply this framework for describing CME in heavy ion collisions~\cite{Hirono:2014oda,Yin:2015fca}. 

\begin{figure*}[t]
\begin{center} 
\includegraphics[scale=0.26]{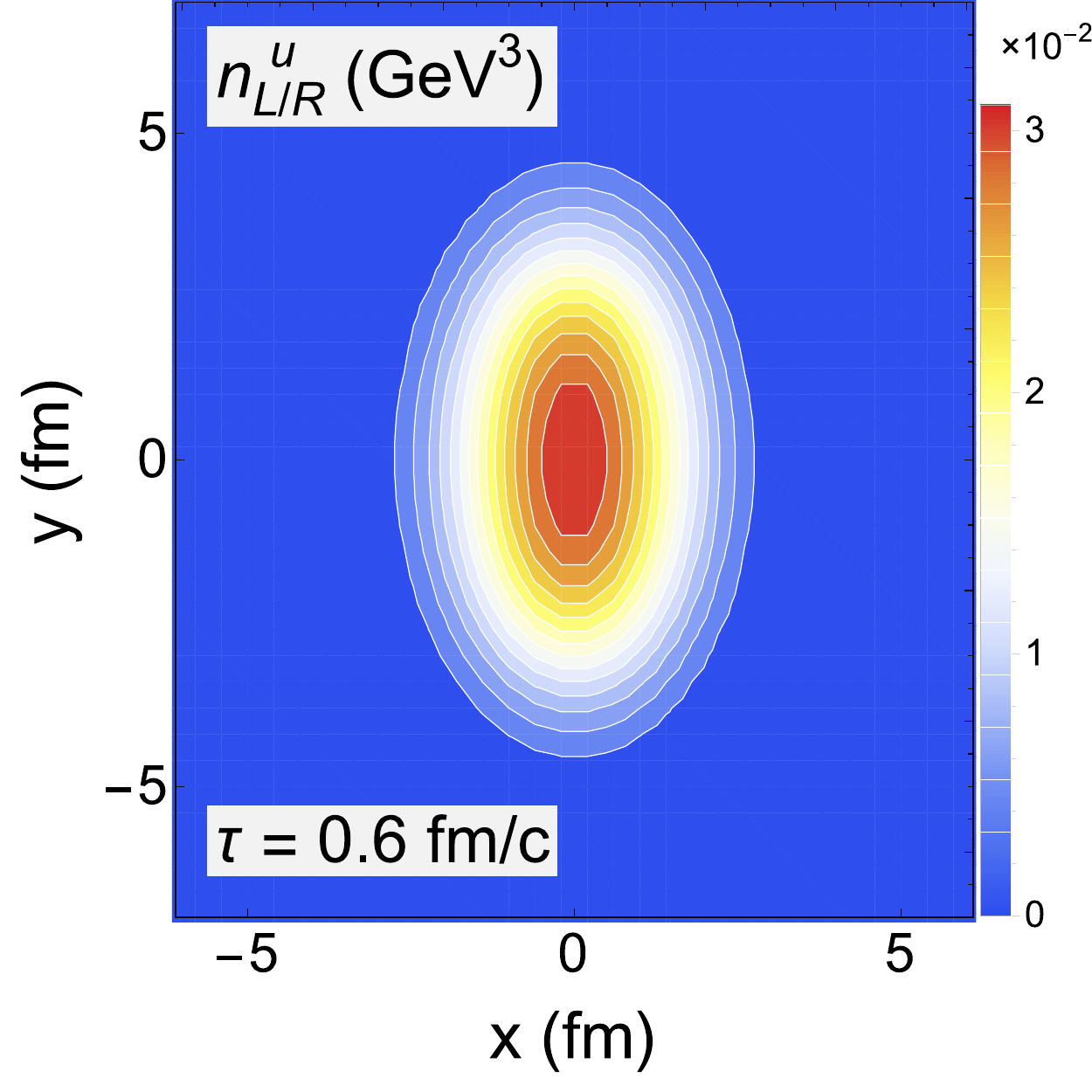} 
\includegraphics[scale=0.26]{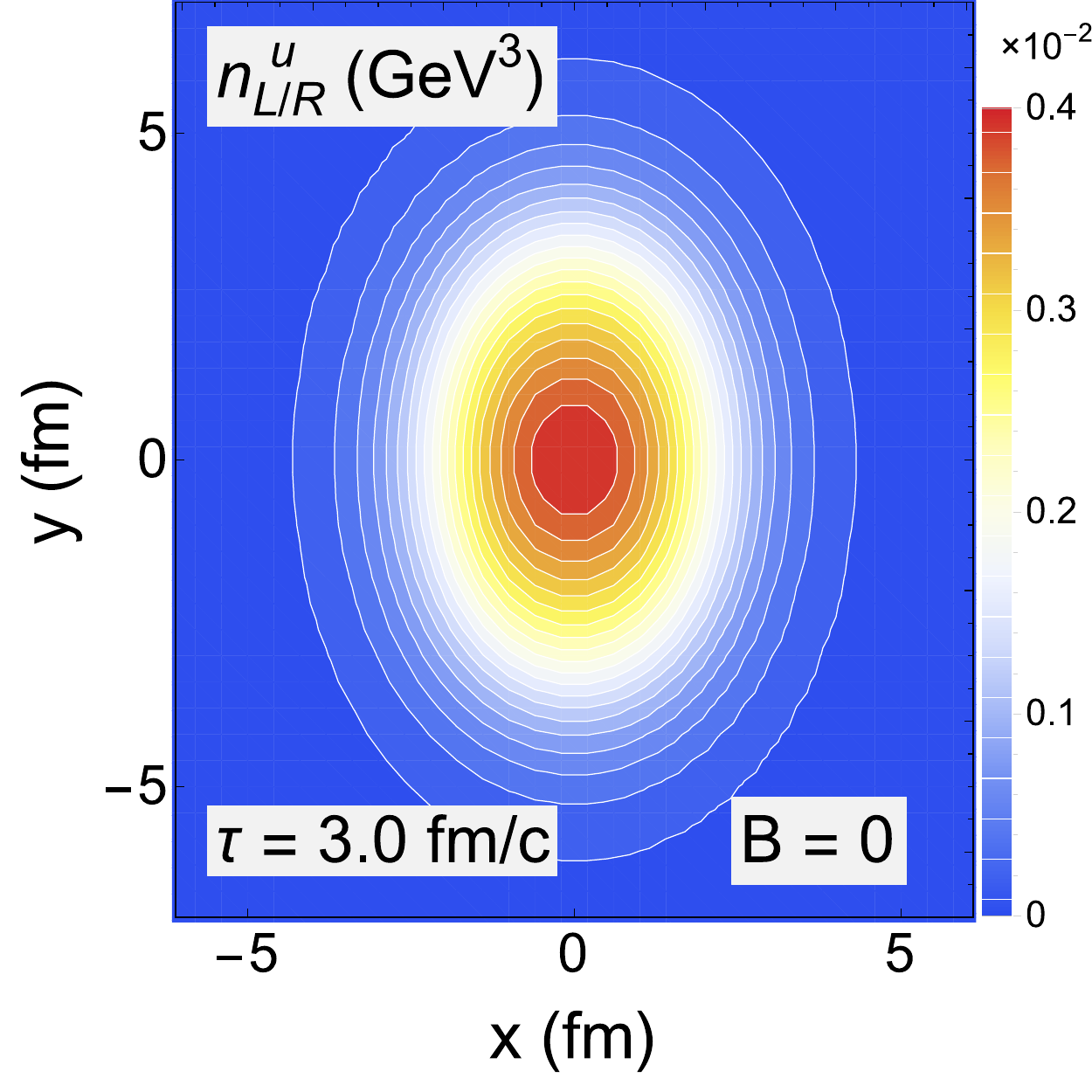}
\includegraphics[scale=0.26]{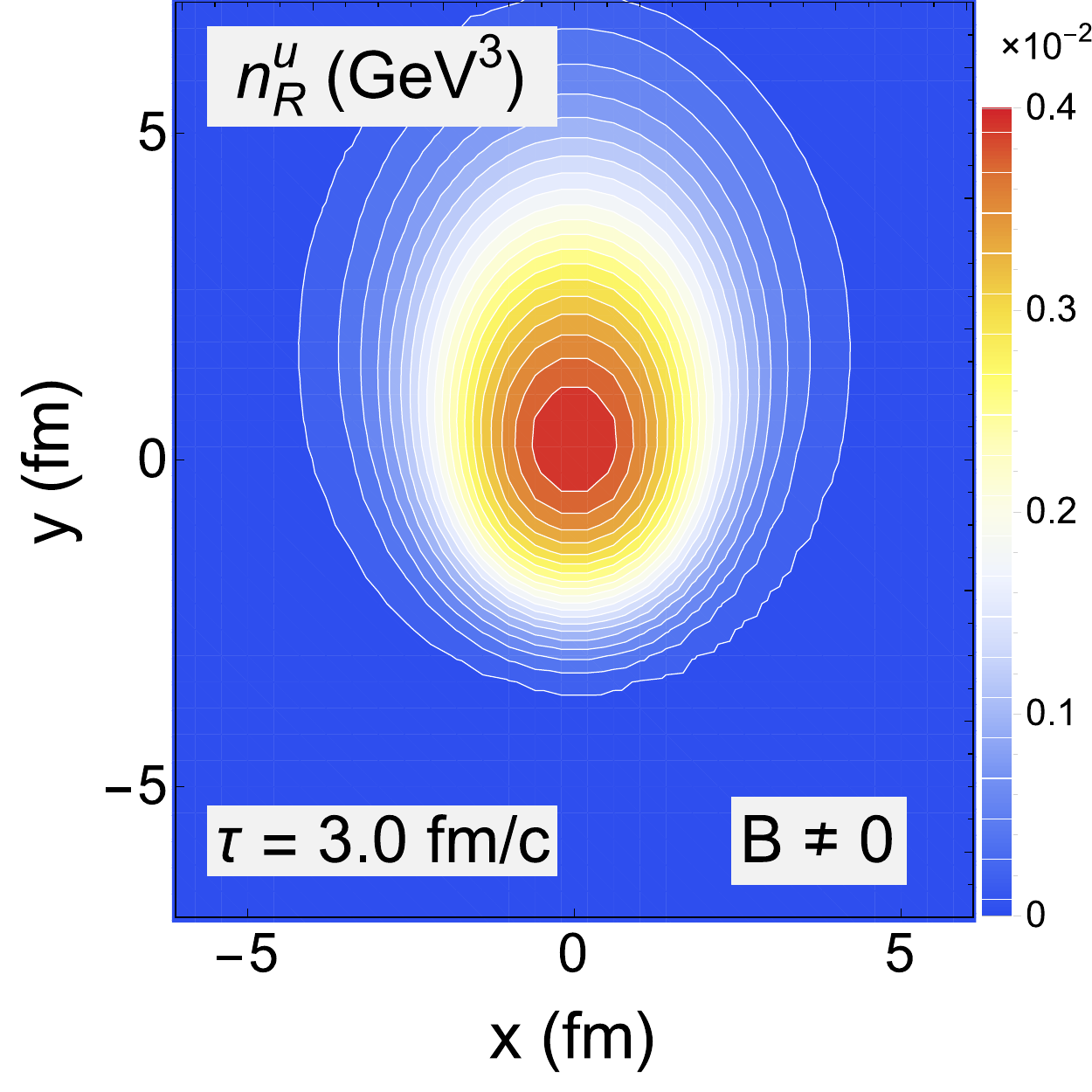} 
\includegraphics[scale=0.26]{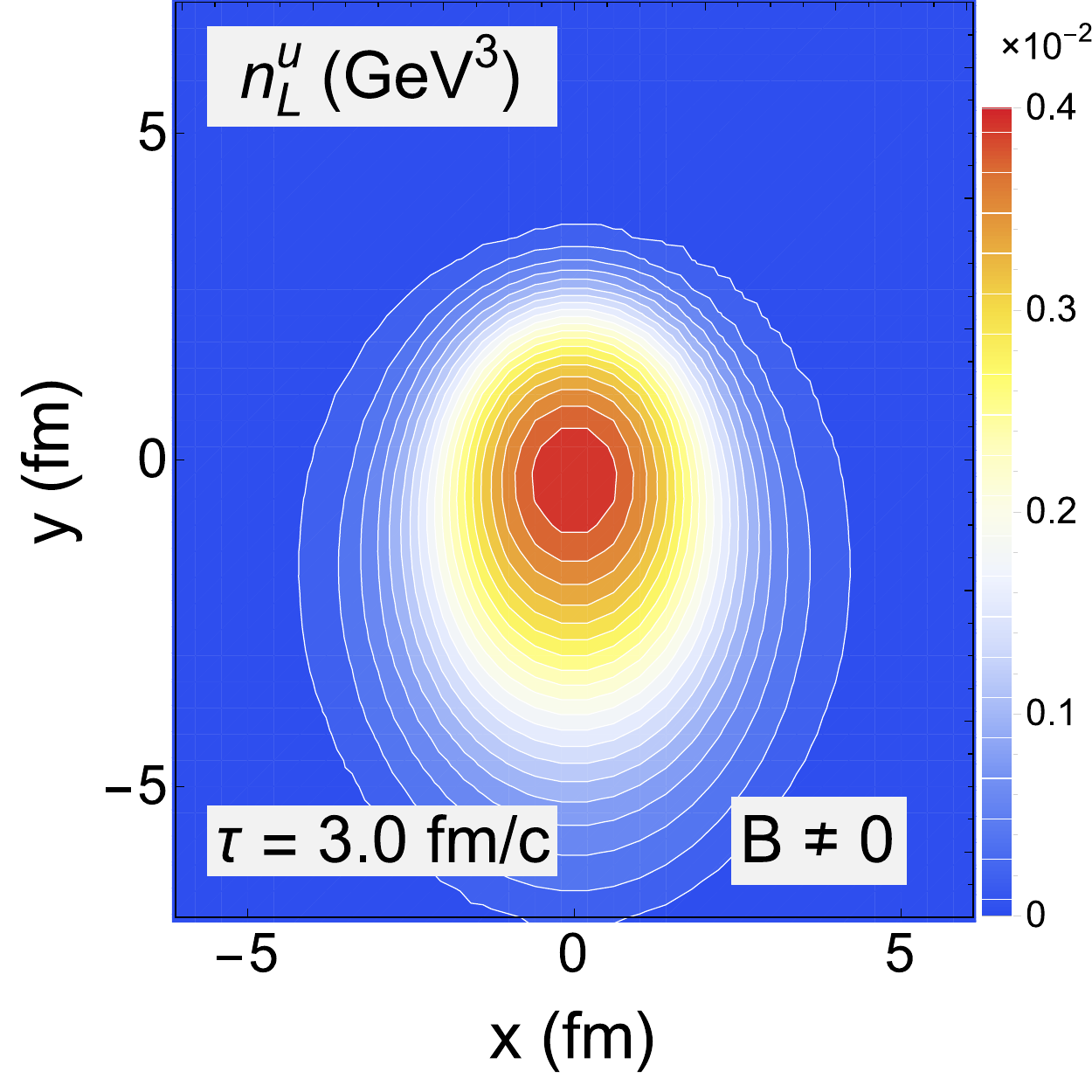}
\caption{(color online) The evolution of u-flavor densities via solving AVFD equations from the same initial charge density distribution (for either RH or LH) at $\tau_0=0.60\rm fm/c$ (left most panel)   in three  cases: (a) (second left panel) for either RH or LH density at $\tau=3.00\rm fm/c$ with magnetic field $B\to 0$ i.e. no anomalous transport; (b) (second right panel) for RH density and (c)  (most right panel) for LH density, both at $\tau=3.00\rm fm/c$ with nonzero $B$ field along positive y-axis. } \label{fig1} \vspace{-0.3in}
\end{center}
\end{figure*}

More recently a sophisticated simulation framework, the Anomalous Viscous Fluid Dynamics (AVFD)~\cite{Jiang:2016wve}, has been developed to describe anomalous chiral transport in heavy ion collisions. In AVFD the bulk QGP evolution is described by the data-validated VISHNU simulations~\cite{Shen:2014vra}. On top of the bulk the chiral currents for up and down flavor quarks are evolved according to anomalous fluid dynamic equations. Furthermore we treat the normal viscous currents $\nu^\mu_{\chi, f}$ at the second-order of gradient expansion by incorporating relaxation evolution toward Navier-Stocks form, thus in consistency with the background bulk flow also described by the 2nd-order viscous hydrodynamics. In Fig.~\ref{fig1} we demonstrate  the most unique feature of the AVFD  framework which lies in the $B$-field driven anomalous current distinguishing the LH  from the RH currents with opposite sign (--- see figure caption and \cite{Jiang:2016wve} for details). 
From the comparison it is evident that in the AVFD framework,  additional transport occurs via anomalous currents along the $B$ field direction, with RH/LH densities evolving in an asymmetric and opposite way.

\section{CME-Induced Charge Separation in Heavy Ion Collisions}

\begin{figure}[!hbt]
\includegraphics[width=2.8in,height=1.8in]{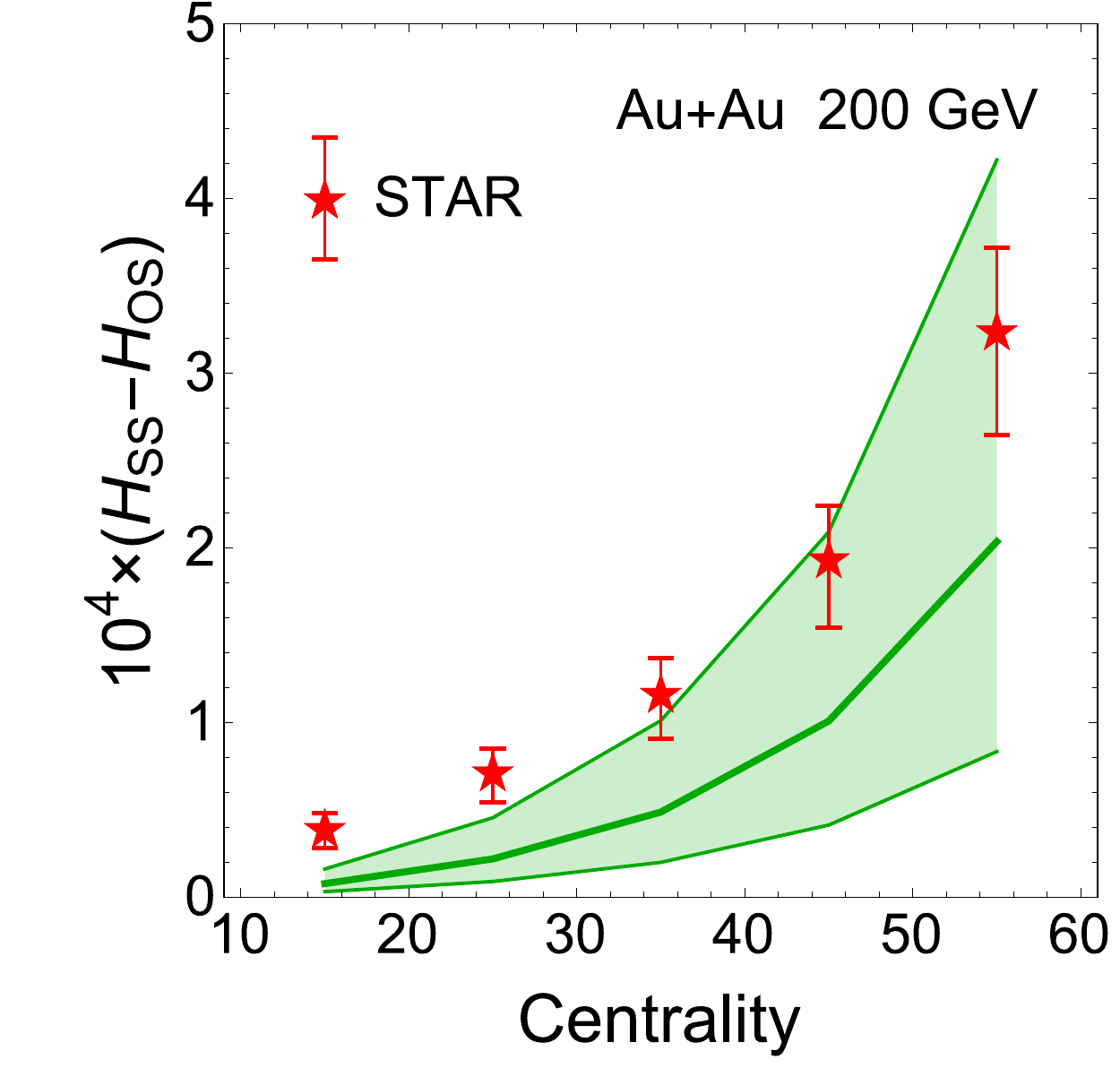} \hspace{2pc}%
\begin{minipage}[b]{14pc}
\begin{center} \vspace{-0.5in}
\caption{\label{fig_H} 
 Quantitative predictions from Anomalous-Viscous Fluid Dynamics simulations for the CME-induced H-correlations, in comparison with STAR measurements~\cite{STAR_LPV_BES}. The green bands reflect current theoretical uncertainty in the initial axial charge generated by gluonic field fluctuations.}
\end{center}
\end{minipage}\vspace{-0.2in}
\end{figure}

In heavy ion collisions, there is very strong magnetic field associated with the fast moving ions at the early stage. Such $\vec B$ field points approximately in the direction perpendicular to the collision reaction plane. The CME current (\ref{eq_CME}) thus leads to charge transport along the $\vec B$ direction and induces a separation of electric charges across the reaction plane. Such a charge separation leads to a dipole term in the azimuthal distribution of produced charged hadrons: 
$\frac{dN^{ch}}{d\phi} \propto [1\pm 2 a^{ch}_1 \sin\phi   + ...]$ 
where $\phi$ is the azimuthal angle measured with respect to the reaction plane, and the $\pm a^{ch}_1$ for  opposite charges respectively. Experimentally such a dipole can be measured through suitable charged-particle-pair azimuthal correlations~\cite{STAR_LPV1,STAR_LPV_BES,ALICE_LPV}. A most pressing challenge in such measurements is the contamination by background correlations, and  crucial for reaching a definitive conclusion is the quantitative prediction for CME-induced contributions to be compared with data. The measurements could be improved by subtracting the bulk-flow-dependent part in the correlations~\cite{Bzdak:2012ia} and the resulting so-called H-correlations, measured by STAR~\cite{STAR_LPV_BES}, could be considered as our ``best guess'' for CME signal at the moment. 
 With the AVFD simulation tool introduced above, we are now ready to quantify the CME-induced charge separation signal under realistic conditions in heavy ion collisions. The magnetic field is from simulations in \cite{Bloczynski:2012en} and the initial chirality imbalance is estimated from initial gluonic field fluctuations~\cite{Hirono:2014oda,Mace:2016svc}. In Fig.~\ref{fig_H} the AVFD predictions are compared with STAR data. Subject to current uncertainties, the AVFD-predicted CME signal with realistic initial conditions and magnetic field lifetime is quantitatively consistent with measurements from 200AGeV AuAu collisions at RHIC. 

\begin{figure*}[t]
\begin{center} \label{fig_iso}
\includegraphics[scale=0.36]{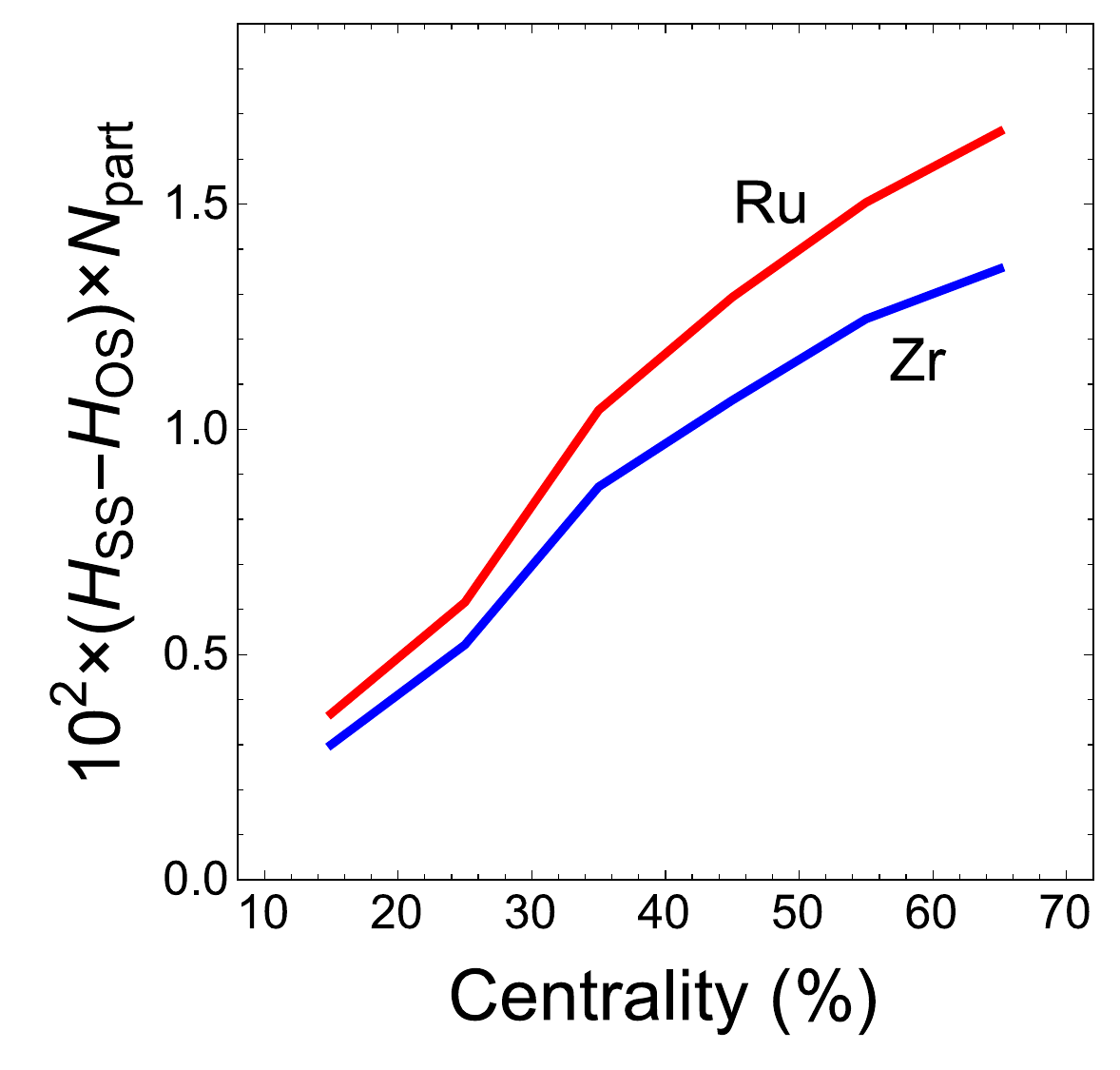} \hspace{0.1in}
\includegraphics[scale=0.39]{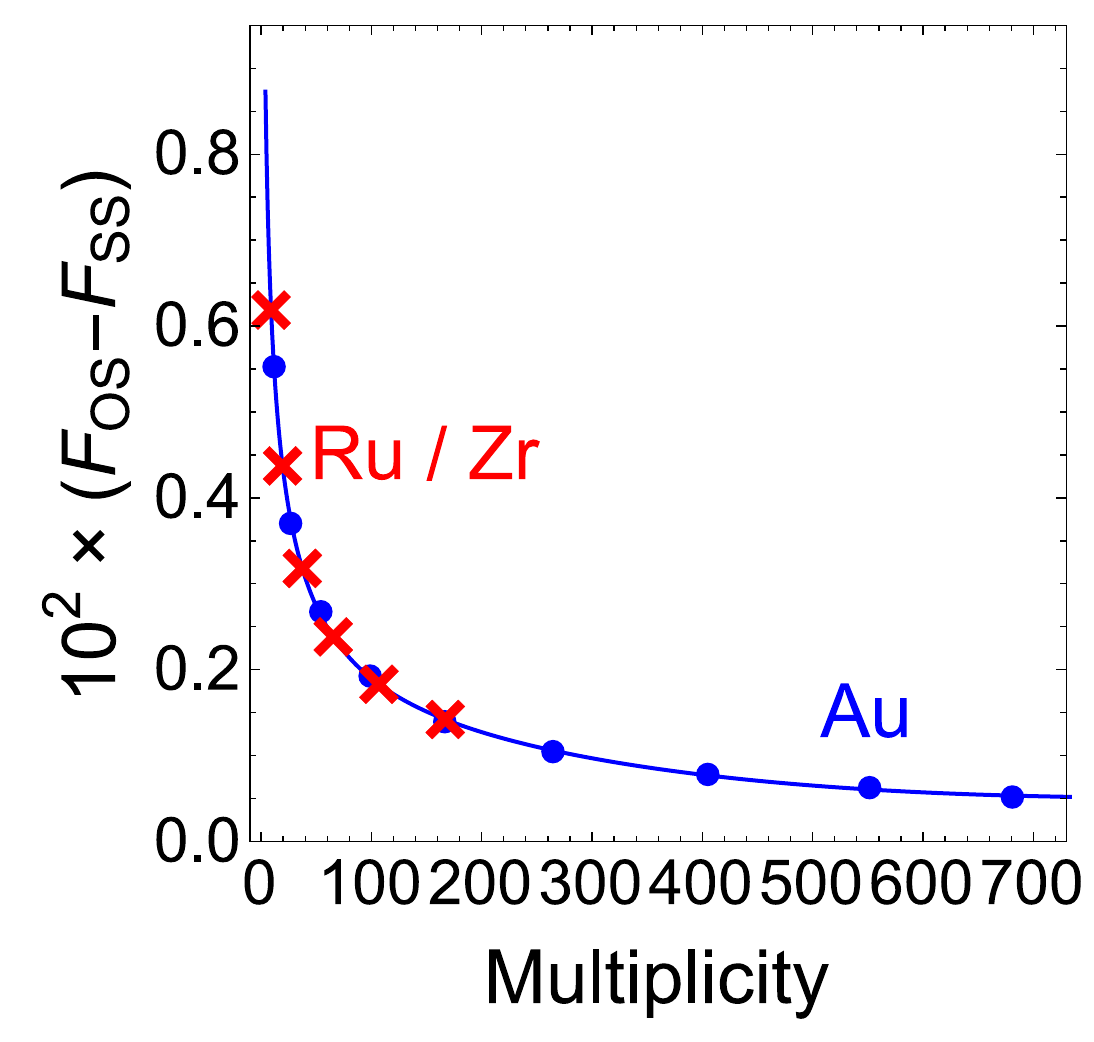} \hspace{0.1in}
\includegraphics[scale=0.36]{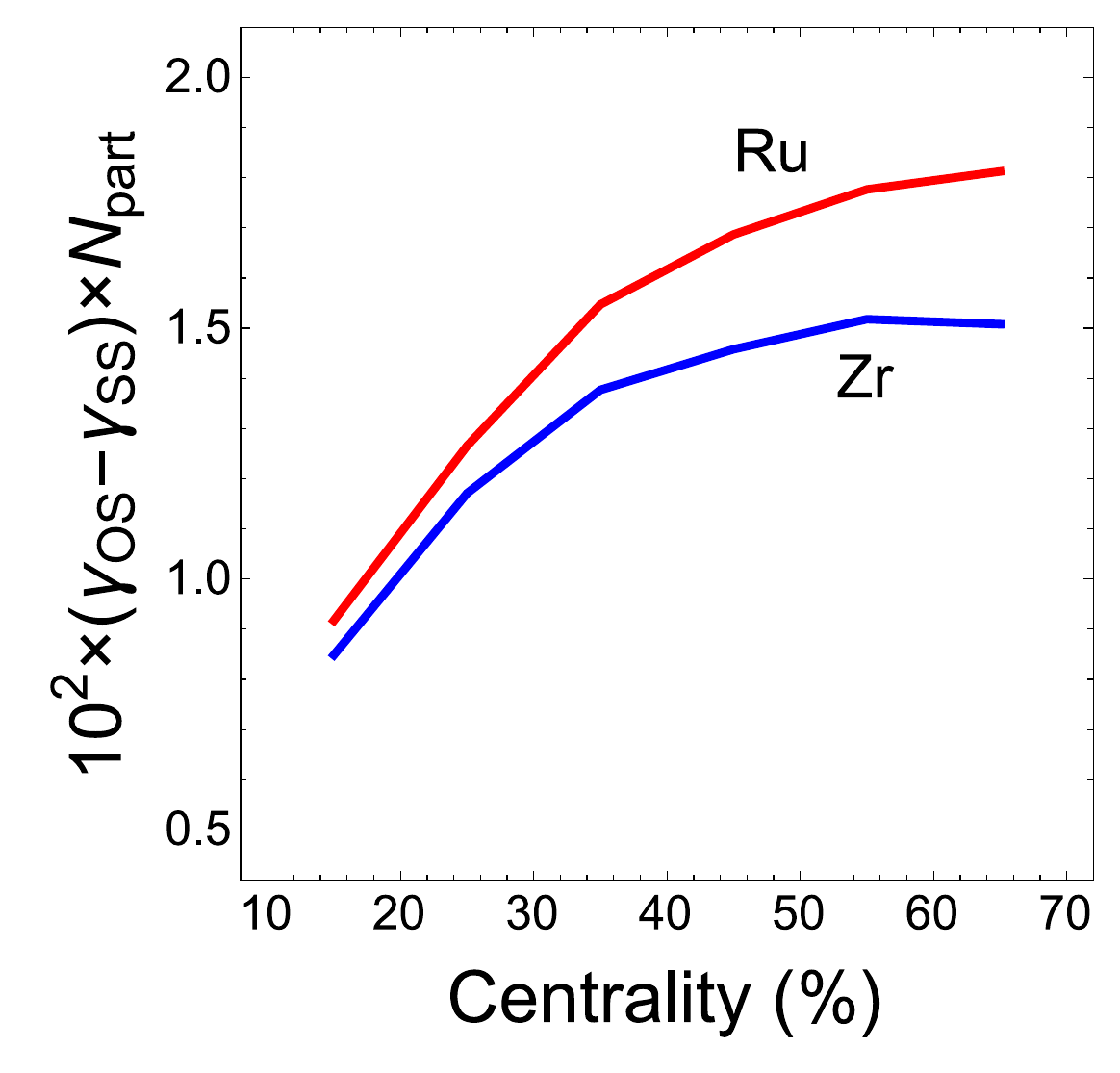}
\caption{(color online) (left) AVFD predictions for CME-induced H-correlations in isobar collisions; (middle) Extrapolation of background-induced F-correlations from AuAu to ZrZr and RuRu systems; (right) Predicted $\gamma$-correlations in ZrZr and RuRu collisions by folding together F and H correlations. }  \vspace{-0.3in}
\end{center}
\end{figure*}

To unambiguously decipher CME from backgrounds, the isobaric collision experiment has been planned at RHIC~\cite{Skokov:2016yrj}. In such ``contrast'' colliding systems (specifically for ZrZr versus RuRu), they have the same bulk evolutions while differ in their magnetic field strength by about $10\%$: therefore a shift of about $20\%$ should be expected for CME-driven correlations on top of identical backgrounds between the two. This will be a crucial  test for the search of CME, and quantitative predictions are important. In Fig.~\ref{fig_iso} we show the AVFD results for the correlations in isobar collisions, with a visible $\sim 15\%$ shift in peripheral region differentiating the two systems.  

\begin{figure}
\includegraphics[width=2.6in,height=1.6in]{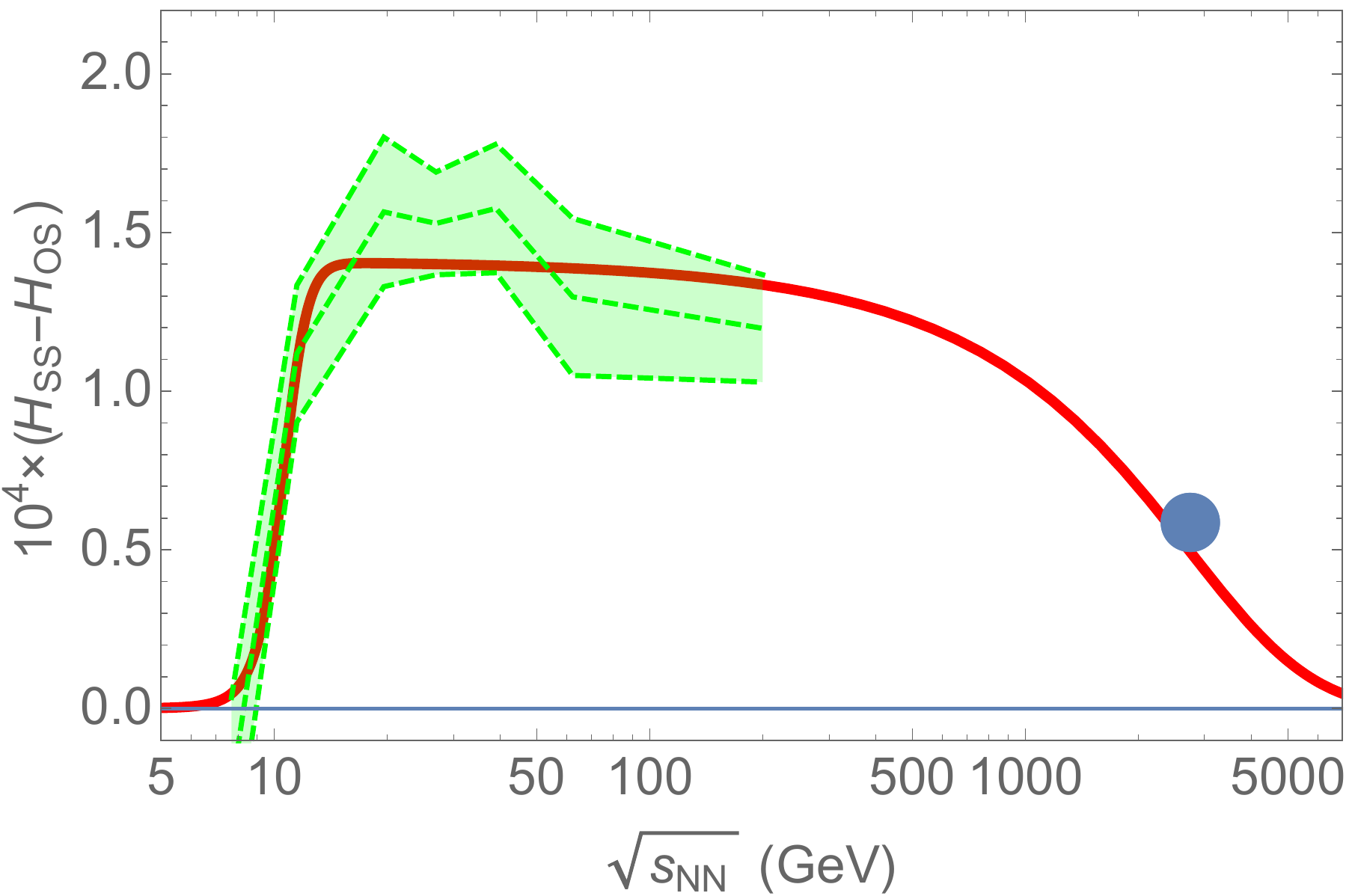} \hspace{2pc}%
\begin{minipage}[b]{14pc}
\begin{center} \vspace{-0.4in}
\caption{\label{fig_E}  (Color online) 
  Fitting the dependence of H-correlations as CME-signal on the collisional beam energy $\sqrt{s}$ based on a simple formula (see text for details). The green band is from RHIC BES data~\cite{STAR_LPV_BES} and the big blue dot is from ALICE data~\cite{ALICE_LPV}, while the red solid curve is from the fitting result. }
\end{center}
\end{minipage}\vspace{-0.3in}
\end{figure}

One interesting question is how the CME signal may change with varying collisional beam energy. The CME relies upon the existence of a chirally-restored QGP phase. Therefore at low enough beam energy where a QGP ceases to form, the CME should disappear. On the other hand the CME requires the magnetic field which however has its lifetime duration decrease with increasing beam energy. At certain high enough beam energy the B field will last for so short a time that no quarks are borne out yet. In that case there will be no CME either. In short, the CME shall ``turn off'' at both very low and very high beam energy, which entertains a rather simple formula for the CME-signal dependence on beam energy $\sqrt{s}$: $H \propto \left[ 1+ \tanh[(\sqrt{s}-\sqrt{s_l})/\Delta_l]\right] \, \left[ 1+ \tanh[(\sqrt{s_h}-\sqrt{s})/\Delta_h]\right]$. In Fig.~\ref{fig_E} we show the fitting result with this formula for RHIC BES data~\cite{STAR_LPV_BES} plus ALICE data point~\cite{ALICE_LPV}. Despite its oversimplification, a quite reasonable fit can be obtained, indicating that it might capture the essential features of the beam energy dependence. The trend of the fitting curve also implies that at LHC $\sqrt{s}=5.05\rm TeV$ collisions, the CME signal may already become negligibly small. Given this perspective, the recently reported CMS measurements at LHC $\sqrt{s}=5.05\rm TeV$ collisions~\cite{Khachatryan:2016got} (which shows a comparable $\gamma$-correlations between $pPb$ and $PbPb$ at the same multiplicity and implies dominance of backgrounds) may not be that surprising after all.

\section{Summary}

In summary we've discussed the manifestation of microscopic chiral anomaly as macroscopic anomalous chiral transport in materials with chiral fermions, focusing on the Chiral Magnetic Effect. In particular reported  the latest theoretical and experimental status in the search of CME in heavy ion collision experiments. Predictions for CME-induced charge separation signal from recently built AVFD simulation framework with realistic initial conditions and magnetic field lifetime are quantitatively consistent with RHIC data. Quantitative  predictions for isobaric collisions and discussions on the disappearance of CME at very low/high beam energies are also presented. We end by briefly mentioning a few other interesting examples of anomalous chiral transport effects: 1) In external magnetic field, vector and axial charge fluctuations are predicted to entangle together and form a propagating wave called the Chiral Magnetic Wave, the signal of which has been confirmed by experimental data~\cite{Burnier:2011bf}; 2) The QGP in non-central collisions bears nonzero fluid rotation with finite vorticity that induces similar anomalous transport as the magnetic field~\cite{Kharzeev:2010gr}; 3) Strong electric field could also induce interesting separation of axial charge which might become observable in asymmetric (e.g. CuAu) collisions~\cite{Huang:2013iia}.

\end{document}